\documentclass[conference]{IEEEtran}

\usepackage{epstopdf}

\usepackage[normalem]{ulem}

\usepackage{times}
\usepackage{datetime}
\usepackage{url}
\usepackage{pdfpages}
\usepackage{amsmath,comment}
\usepackage{epsfig,multirow,color}
\usepackage{algorithm}
\usepackage{algorithmicx}
\usepackage{algpseudocode}
\usepackage{wrapfig}
\usepackage{comment}
\usepackage{amsthm}
\usepackage{url}
\usepackage{multirow}
\usepackage{graphicx}
\usepackage{xspace}
\usepackage{textcomp}
\usepackage{listings}
\usepackage{todonotes}
\usepackage{layouts}

\usepackage{booktabs} 
\usepackage{graphicx}
\usepackage{subcaption} 
\usepackage[numbers]{natbib}
\usepackage{balance}

\algnewcommand\algorithmicinput{\textbf{INPUT:}}
\algnewcommand\algorithmicoutput{\textbf{OUTPUT:}}
\algnewcommand\INPUT{\item[\algorithmicinput]}
\algnewcommand\OUTPUT{\item[\algorithmicoutput]}

\newcommand{\code}[1]{\texttt{{\small #1}}}
\newcommand{\ignore}[1]{}

\usepackage{color}
\usepackage{colortbl}
\usepackage{ulem}

\definecolor{mygreen}{rgb}{0,0.4,0}
\definecolor{mygray}{rgb}{0.5,0.5,0.5}
\definecolor{mymauve}{rgb}{0.58,0,0.82}

\usepackage{titling}
\usepackage{hyperref}
\hypersetup{
    colorlinks=true,
    linkcolor=blue,
    citecolor=blue,
    urlcolor=blue,
    filecolor=blue,
    bookmarks=true,
}
\usepackage{cleveref}
\usepackage[affil-it]{authblk}
\crefformat{section}{\S#2#1#3}
\crefname{figure}{Figure}{Figures}
\crefname{table}{Table}{Tables}
\crefname{algorithm}{Algorithm}{Algorithms}

\newcommand{\name}{PATTER}
\newcommand{\tab}{\hspace*{1em}}

\begin{document}
\title{ARM Pointer Authentication based Forward-Edge and Backward-Edge Control Flow Integrity for Kernels}

\date{}
\IEEEoverridecommandlockouts
\author{Yutian Yang}
\author{Songbo Zhu}
\author{Wenbo Shen\thanks{Corresponding author.}}
\author{Yajin Zhou}
\author{Jiadong Sun}
\author{Kui Ren\\ \textit{\{ytyang, 3160103828, shenwenbo, yajin\_zhou, simonsun, kuiren\}@zju.edu.cn}}

\affil[]{Zhejiang University}

\maketitle

\thispagestyle{plain}
\pagestyle{plain}

\thispagestyle{empty}

\begin{abstract}

Code reuse attacks are still big threats to software and system security. Control flow integrity is a promising technique to defend against such attacks. However, its effectiveness has been weakened due to the inaccurate control flow graph and practical strategy to trade security for performance. In recent years, CPU vendors have integrated hardware features as countermeasures. For instance, ARM Pointer Authentication (PA in short) was introduced in ARMV8-A architecture. It can efficiently generate an authentication code for an address, which is encoded in the unused bits of the address. When the address is de-referenced, the authentication code is checked to ensure its integrity. Though there exist systems that adopt PA to harden user programs, how to effectively use PA to protect OS kernels is still an open research question. 

In this paper, we shed lights on how to leverage PA to protect control flows, including function pointers and return addresses, of Linux kernel. Specifically, to protect function pointers, we embed authentication code into them, track their propagation and verify their values when loading from memory or branching to targets. To further defend against the pointer substitution attack, we use the function pointer address as its context, and take a clean design to propagate the address by piggybacking it into the pointer value. We have implemented a prototype system with LLVM to identify function pointers, add authentication code and verify function pointers by emitting new machine instructions. We applied this system to Linux kernel, and solved numerous practical issues, e.g., function pointer comparison and arithmetic operations. The security analysis shows that our system can protect \text{all} function pointers and return addresses in Linux kernel.

\end{abstract}
\IEEEpeerreviewmaketitle

\section{Introduction}
\label{sec:intro}

Since the first emerging in the 1990s~\cite{solar-rop}, code reuse attack has become a big threat to software and system security, especially after code injection has been defeated by hardware features, including NX/SMEP/SMAP on x86 and XN/PXN/PAN on ARM.
Specifically, after hijacking the control flow through memory corruption, attackers could chain existing code snippets (called code gadgets) together to perform malicious operations. This is called return-oriented programming (ROP in short)~\cite{ROP, ROPRISC}. Previous studies showed that given a large codebase (such as Linux kernel or libc), ROP has been shown to be Turing complete~\cite{roemer2012return}, making it a powerful attack.




To defend against the ROP attack, multiple solutions have been proposed, which are roughly falling into two categories.
The first category includes systems to make attackers hard to obtain necessary information to launch the attack, either by randomizing memory layout~\cite{aslr, aslr1, Oxymoron,wartell2012binary}, or reducing the number of available gadgets~\cite{GFREE}. 
However, address randomization has been proven to be ineffective~\cite{aslrattack1,aslrattackT1}, since the address information could be leaked or inferred. Moreover, the large codebase makes it impossible to totally eliminate code gadgets. 
The second category includes systems to protect the integrity of control flow (CFI in short)~\cite{CFI,zhang2013practical,zhang2013control}. Though CFI is a promising technique, its effectiveness
has been weakened~\cite{CFB} due to the inaccurate control flow graph and the practical strategy to trade security for performance.

In recent years, hardware-assisted control flow enforcement~\cite{mashtizadeh2015ccfi,mohan2015opaque} has drawn much attention. These systems mainly borrow hardware features that were designed for other purposes. Nowadays, vendors have directly embedded security features for CFI in modern CPUs. 
For instance, ARM introduced Pointer Authentication (PA) in ARMv8.3~\cite{arm-pa}.
Specifically, it reuses unused bits in the virtual address of the ARM64 architecture to calculates and embed an authentication code for the pointer, thus the name Pointer Authentication Code (PAC). When the pointer is de-referenced, the embedded authentication code could be used to verify its validity by the hardware. To facilitate its use, multiple instructions are added.



Since its debut, PA has been considered as a promising defense due to its powerful security guarantees and efficient pointer value verification~\cite{qualcomm-ret-addr}. However, to leverage this feature, programmers need to change and recompile their programs to use the new instructions. Though a couple of papers are adopting PA to protect code and data pointers in user programs~\cite{hans2019pac, HANSCALLSTACK, HANSCANARY}, there is no open implementations that leverage PA to protect privileged software, i.e., OS kernel~\footnote{We are aware that Apple has adopted PA in its latest version of iOS XNU kernel. However, its implementation details are unknown.}. Due to the differences between OS and user programs (for instance, while user programs could assume that the underlying kernel is trusted to provide cryptography keys to generate the authentication code, OS kernels cannot make such an assumption), how to  effectively use PA to protect OS kernel is still an open research question.



\noindent\textbf{Our work}\tab
In this paper, we shed lights on how to leverage PA to protect control flows of OS kernels, and present the first design and implementation of such a system. 
Specifically, we propose \textbf{PATTER}, which is short for \uline{P}ointer \uline{A}u\uline{T}hen\uline{T}ication for k\uline{ER}nels, protects both function pointers and return addresses in Linux kernel, thus providing both forward- and backward-edge control flow integrity. To the best of our knowledge, it is the first \textit{open} implementation of applying PA to Linux kernel. 




In order to leverage PA to provide complete protection of function pointers, \name{} needs to append the authentication code to the value of a function pointer~\footnote{In this paper, if not specified, we use ``function pointer'' to denote its value, i.e., the jump target.}, track the propagation of function pointers, and verify its validity when loading its value from memory or branching to the jump target.
Specifically, \name{} calculates an authentication code (PAC) for each function pointer (we call the function pointer with an authentication code as a PACed pointer) before it is written into memory. The PAC is computed using the combination of a hardware cryptography key, the function pointer value, and a context. Then \name{} tracks the propagation of the function pointer with the help of the LLVM compiler. When a PACed pointer is loaded from the memory, \name{} verifies the value to ensure that it has not been modified (attackers have arbitrary memory write capability).
However, the previous step is not enough since an attacker could directly jump before the instruction that dereferences a function pointer (the \code{blr} instruction for instance). In this case, \name{} verifies the jump target in indirect branch instructions before jumping to it.

When calculating the authentication code, unlike the previous work that leverages the function type as a context~\cite{hans2019pac}, \name{} takes the address of a function pointer as its context. That's because for each function pointer, the function type is not unique. Attackers could obtain the PACed function pointer and reuse it for another function pointer. This is called pointer substitution attack. By using the unique address of a function pointer as its context, \name{} is immune to this attack.
However, the challenge is the location of de-referencing a function pointer may be far from the location where it is loaded, thus we need to propagate the address of a function pointer between procedures. \name{} takes a clean design to piggyback the pointer address into the pointer value (Figure~\ref{fig:piggyback}) to solve this problem.

To protect return addresses, \name{} will generate the PAC for a return address before saving it to the stack and check the PAC after loading it from the stack.
The stack pointer is used as the context, so that the signed return addresses cannot be replayed across different stack frames. As a result, the return address corruption and return address replay attacks will be defeated by \name{}. Moreover, different from existing works, \name{} uses a single instruction to authenticate the loading return address and return it atomically, which defeats the time of check to time of use attacks.

We have implemented a prototype system with LLVM and applied it to Linux kernel v5.0.1. Specifically, we developed LLVM passes to identify function pointers, add authentication code, propagate and verify function pointers by emitting new machine instructions into the binary. To apply \name{} to Linux kernel, we also modified the kernel to patch the statically initialized function pointers, and solved multiple practical issues, including function pointer comparison, function pointer arithmetic operations and function pointers inside a union. 
The security analysis shows that \name{} can protect \textit{all} the function pointers and return addresses in Linux kernel, with a performance overhead between $15$\% to $25\%$, using the micro benchmark of system calls.

This paper has the following contributions:
\begin{itemize}
    \item  We propose the first design of using the ARM pointer authentication to protect control flow transfers of Linux kernels. Our design protects \textit{all} function pointers and return addresses in Linux kernel, thus providing both forward-edge and backward-edge control flow integrity. 
    
     
    
    
    \item To implement \name{}, we have proposed a series of new techniques to solve technical challenges. In particular, we proposed \textit{address-base} authentication code generation to defend against pointer substitution attacks, and 
    \textit{pointer address piggyback} to propagate function pointer address. We also proposed methods to identify function pointers, and verify them when loading, storing their values, and branching into targets.
    
    
    \item We have implemented a prototype of \name{} based on the latest Clang/LLVM and applied it to protect the latest version of the Linux kernel. \name{} successfully protects 100\% of indirect call sites and return addresses.
    

\end{itemize}

The organization of this paper is as follows:  background knowledge is given in~\cref{sec:background}. \cref{sec:threat} discusses the threat model and assumptions. \name{} design is presented in detail in~\ref{sec:design}. We discuss the implementation details in~\cref{sec:imple} and evaluate both the security and performance of \name{} in~\cref{sec:eval}. We compare \name{} with related works in~\cref{sec:related}. Finally, we conclude the whole paper in~\cref{sec:conclu}.

\begin{figure}
    \centering
    \includegraphics[width=\linewidth]{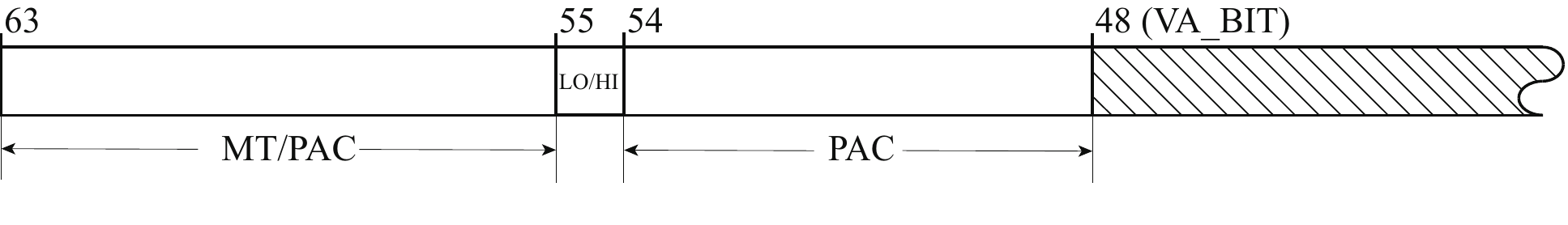}
    \caption{ARMv8.3 Pointer Format with Pointer Authentication. The Pointer Authentication Code (PAC) is embedded into the unused bits of a pointer.}
    \label{fig:general_PAC}
    \vspace{-3ex}
\end{figure}

\section{Background}
\label{sec:background}

In this section, we give preliminary background knowledge of the techniques used by this paper, including pointer authentication and ROP/JOP attacks.

\subsection{ARMv8.3 Pointer Authentication}
\label{sec:background-pa}

ARM has introduced a new hardware security feature in ARMv8.3, named Pointer Authentication (PA)~\cite{ARM_PA}, to protect integrity of pointers saved in memory. The basic idea of PA is to compute a cryptographic keyed hash, trunk the hash and embed it into the unused bits in the pointer. The functionality of the cryptographic keyed hash is the same to message authentication code (MAC), therefore it is termed as Pointer Authentication Code (PAC).

Figure \ref{fig:general_PAC} depicts the format of a PACed pointer on ARM64. \code{VA\_BIT} represents the size of the virtual address space, which is usually 39 or 48 bits. The other bits of a pointer are not used for address translation, therefore can be used to hold the PAC. Note that the top 8 bits are occupied by memory tag if the ARM memory tag extension (MTE)~\cite{ARM_MTE} is enabled. Therefore, depending on the configuration, the PAC size can be 7 or 15 bits.
ARMv8.3 PA uses QARMA block cipher algorithm~\cite{avanzi2017qarma} for PAC generation:
\begin{center}
\code{PAC = QARMA(key, pointer, context)}.
\label{eq:pac}
\end{center}

QARMA takes a 64-bit pointer with a 64-bit context as inputs and outputs a  64-bit cipher block. QARMA uses 128-bit key, which is kept in dedicated registers. ARMv8.3 PA provides five key registers, out of which,  \code{APIAKey} and \code{APIBKey} are designed for encrypting code pointers; \code{APDAKey} and \code{APDBKey} are designed for encrypting data pointers; and  \code{APGAKey} can be used for general purpose. The output cipher is then truncated to a suitable size and embedded into the PAC field showed in Figure\ref{fig:general_PAC}.  

ARMv8.3 also provides a new set of instructions for PA support. \code{pac*} instruction are designed for generating and embedding the PAC. For example, \code{pacia x0, x1} accepts \code{x0} as the pointer and \code{x1} as the context, generates the PAC using \code{APIAKey}, and embeds the PAC into \code{x0}. Correspondingly, \code{aut*} instructions are designed for PAC authentication. For example, \code{autia x0, x1} will verify the PAC embedded in \code{x0} by using \code{APIAKey} and \code{x1} as the context, if \code{x0} has a valid PAC, x0 will be changed into a normal pointer, otherwise its top bits will be flipped and an address translation error will be triggered upon de-referencing the pointer. 

Some instructions are designed for specific usages, such as \code{paciasp} generates the PAC using \code{x30} as the pointer and stack pointer \code{sp} as the context by default. Similarly, \code{autiasp} authenticates the PAC using \code{x30} as the pointer and stack pointer \code{sp} as the context.

Besides the basic PAC generation and authentication instructions, ARMv8.3 PA also provides PA combined instructions. For example, function pointer branch (call) operation is usually done by \code{blr} instruction, which branches to the function pointer and updates the link register with the correct return address. PA now provides \code{blraa}, which authenticates the function pointer first before branching to function pointer. Similarly, for the function return, PA provides \code{retaa}, which authenticates the return address before returning to it.

Guarded by pointer authentication, even though the attacker can corrupt function pointers with memory corruption vulnerabilities, the corrupted function pointer cannot pass the authentication without the knowledge of the key. Therefore, ARMv8.3 PA can be a cornerstone for designing new control flow protection schemes.
However, due to the limited number of key registers, PA is vulnerable to pointer substitution attacks if the context is not selected properly. Existing works~\cite{hans2019pac} propose to use function types as the context, which still allows the same type function pointer substitution attacks.


\subsection{ROP and JOP Attacks}

Software memory corruption bugs have existed for more than 30 years~\cite{Song13oakland}, during which lots of attacks and defenses mechanisms have been proposed. In early years, attackers would inject assembly codes (called \textit{shellcode}) into the application memory and then jump to the injected instructions. However, since Data Execution Prevention (DEP) has been proposed, the W$\oplus$X has been supported by almost all mainstream architectures, and injection code in the writable memory area became impossible. 

With code injection being defeated, attackers cannot inject new code, begin to reuse existing code to construct new attack functions. This kind of attack is termed as \textit{code reuse attacks}. To launch a code reuse attack, the attacker first hijacks the program's control flow to execute deliberately selected assembly instruction snippets, called \textit{gadgets}. Gadgets can be chained together to construct new functions for malicious ends. 

Depending on the control data that the attacker hijacks, code reuse attacks can be divided into two categories: return-oriented programming (ROP) and jump-oriented programming (JOP).
In return-oriented programming (ROP) attacks, the attacker controls the call stack through vulnerabilities, such as buffer overflows, then by injecting the gadgets' addresses as the return addresses, the control flow of the program is redirected to the gadgets. In ROP, each gadget ends with a return instruction \code{ret}, and that is why it is called return-oriented programming. To defend against ROP attacks, security researchers proposed address space layout randomization (ASLR), which randomizes the address of code, stack, and heap, making it hard to predict the code gadgets' addresses and the buffer overflowed address.  However, ASLR is vulnerable to address leaks~\cite{belleville2019kald}. Its design cannot solve the return address corruption problem fundamentally.

JOP attack, on the other hand, overwrites the function pointers. When the program calls the corrupted function pointer via instructions like \code{blr} or \code{br}, the program’s control flow is hijacked by attackers. Gadgets in JOP end with a jump instruction, such as \code{blr} or \code{br} on ARM and \code{jmp} on x86, hence gains the name of jump-oriented programming.  To defend against the JOP attack, researchers proposed control-flow integrity. The main idea is to check the function pointers jumping targets according to the Control Flow Graph (CFG) or function pointer type so that only the targets are in the CFG or jump targets are the same type with the function pointer, then the jump is allowed.  

\section{Threat Model and Assumptions}
\label{sec:threat}


\subsection{Threat Model}
The attacker in our paper is powerful with arbitrary kernel memory read and write capability. However, the attacker cannot change existing kernel code or inject new code to the kernel. This is reasonable as W$\oplus$X is supported by all mainstream CPU architectures. Moreover, new isolation based designs~\cite{azab2014hypervision, azab2016skee} use trust execution environments, such as TrustZone~\cite{arm-tz}, to protect the kernel code against different kinds of kernel vulnerabilities.

Even though the attacker gains arbitrary kernel memory read and write capability via kernel vulnerabilities~\cite{zhang2019pex}, he/she still cannot read or write the registers directly, such as the PA key registers. However, the attacker is able to read and write the register contents that are saved to the kernel memory.

With the arbitrary kernel memory read and write capability, the attacker tries his/her best to change the control data, such as function pointers or return addresses, to gain code execution capability in kernel space. The attacker can corrupt the function pointers in the kernel data section, stack, and heap or the return addresses on the kernel stack. The attacker may also try to guess the PAC value or launch pointer substitution attacks to replace the original pointer value with the interested PACed pointer values.

\subsection{Assumptions}
We assume that the kernel boot-up process is trusted. This is a valid assumption as the bootloader can verify the cryptographic hash of the kernel binary easily when loading kernel image to memory. The boot time verification guarantees the integrity of the kernel image, as well as the trustworthiness of the kernel boot-up. After the kernel fully boots up, allowing system calls, that is when the vulnerabilities can be triggered, and the kernel can be attacked.

We further assume the random number generation in the kernel is trusted so that the generated random number has the expected random entropy. As a result, the attacker cannot guess the PA key easily.
%
Finally, we assume that the hardware works as defined by the ARM specification, especially for the Pointer Authentication related hardware.

\begin{figure}
    \centering
    \includegraphics[width=\linewidth]{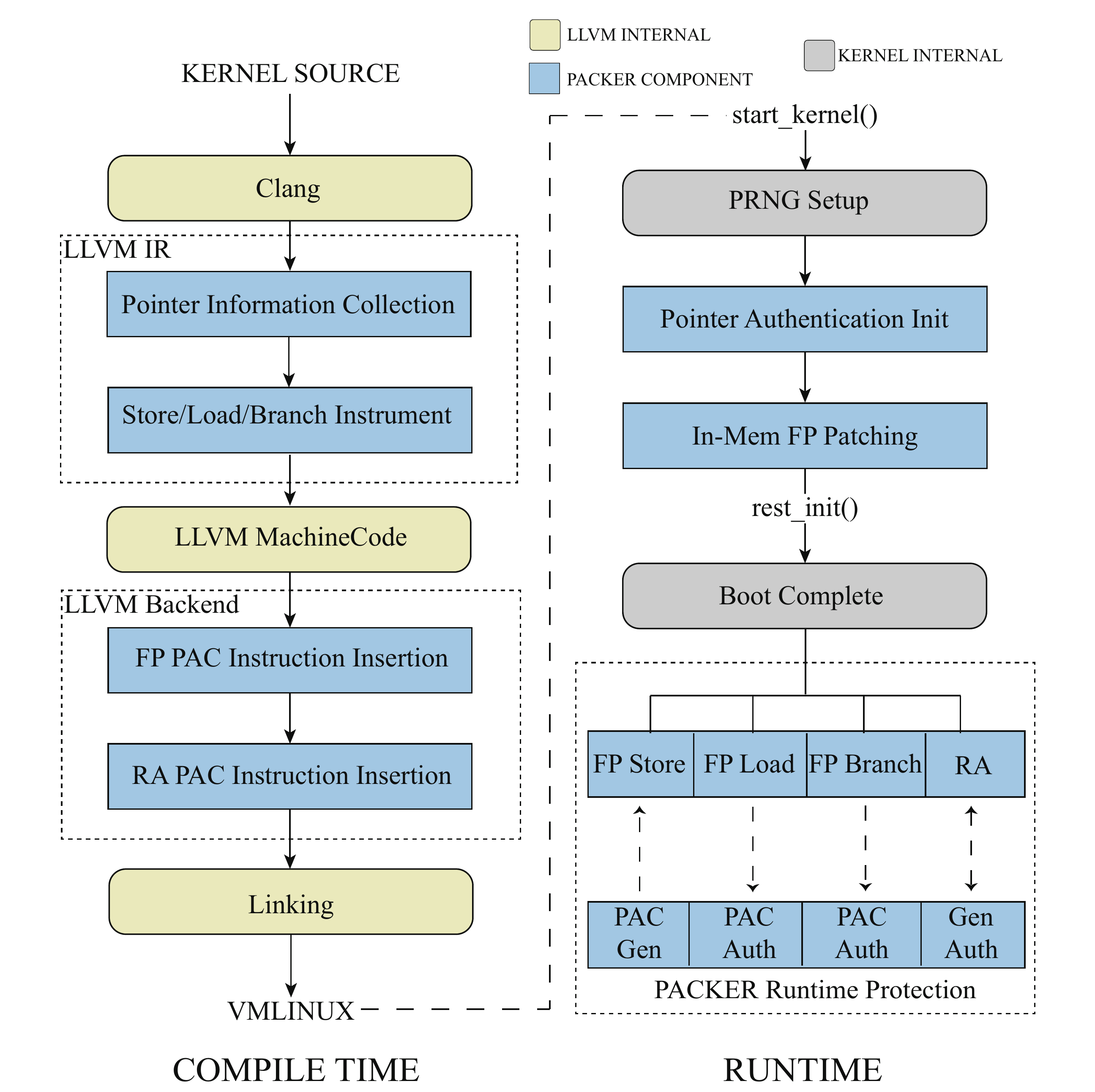}
    \caption{\name{} Overview. \name{} contains two stages: compile time and runtime.}
    \label{fig:overview}
    \vspace{-3ex}
\end{figure}

\section{\name{} Design}
\label{sec:design}

\subsection{Overview}

As mentioned in~\cref{sec:background-pa}, due to the limited number of key registers, ARMv8.3 PA is vulnerable to pointer substitution attacks. For example, even though the attacker does not have the key, it can trigger vulnerabilities to leak a PACed code pointer, and substitute the attacking pointer with this leaked code pointer. The leaked code pointer already has a valid PAC, thus can pass the PA authentication. In this way, the attacker can still launch JOP attacks, in which the function pointers become the replayed PACed function pointers. Existing works~\cite{hans2019pac} proposed to use function type as the context when generating PAC, which achieves the same protection with the fine-grained CFI~\cite{tice2014enforcing} that only allows a function pointer to jump to a set of functions with the same function signature at runtime~\cite{qualcomm-ret-addr}. Pointer substitution attacks still exist for the code pointer with the same type (function signature).

%


To defeat pointer substitution attacks, \name{} proposes \textit{address-based PAC}, in which the virtual address of a function pointer variable is used as its context when computing the PAC. Therefore, we have 
\begin{center}
\code{PAC = QARMA(key, pointer, address)},
\end{center}
where the \code{key} the 128-bit encryption key, the \code{pointer} is the function pointer value while \code{address} is the virtual address of the function pointer variable. 
The basic idea behind address-based PAC is that \textit{all function pointers in kernel memory are within the same address space, therefore all of them have different virtual addresses}. To defend against pointer substitution attacks, \name{} leverages the unique virtual address to generate unique pointer PAC, so that \textit{one PAC is bonded to one particular address, cannot be replayed to other addresses.}

Overall, \name{} consists of two stages: compiling stage and run-time stage, as shown in~\cref{fig:overview}. During the compiling stage, \name{} relies on Clang to compile kernel source code to LLVM IR. Then on the kernel IR, \name{} first analyses global variables and all data structures inside a module, and then identifies function pointers inside the module~\cref{sec:identFP}. 

After that, with the identified function pointer information, \name{} instruments the IR and the backend instructions that involve the function pointer store, load and branch operations~\cref{sec:instrument}. Finally, for the return address, \name{} inserts PAC generation code in the function prologue, as well as the PAC checking code in the function epilogue~\cref{sec:ret-addr-prot}.

After the compiling stage, a PA-instrumented vmlinux binary is generated. During the runtime stage, especially the kernel boot-up process, \name{} first configures the registers and initializes the PA keys~\cref{sec:pa-init}. Then \name{} generates PAC for the statically initialized function pointers and dynamically function pointer assignments that happen before PA initialization~\cref{sec:ptr-init}. After that, \name{} functionality is complete, it protects all code pointers inside kernel memory, including function pointers and return addresses.


\begin{figure}
    \centering
    \includegraphics[width=\linewidth]{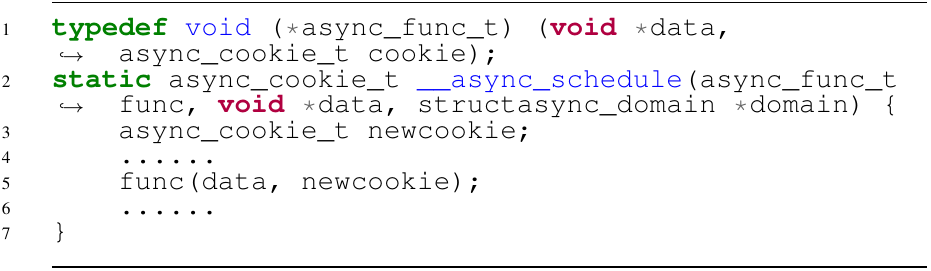}
    \caption{Code in kernel which passes FP as a function argument. The argument \code{func} of \code{\_\_async\_schedule} function is a function pointer, its address information is lost at its call site in Line 5.}
    \label{fig:FPArg}
    \vspace{-3ex}
\end{figure}

\begin{figure}
    \centering
    \includegraphics[width=\linewidth]{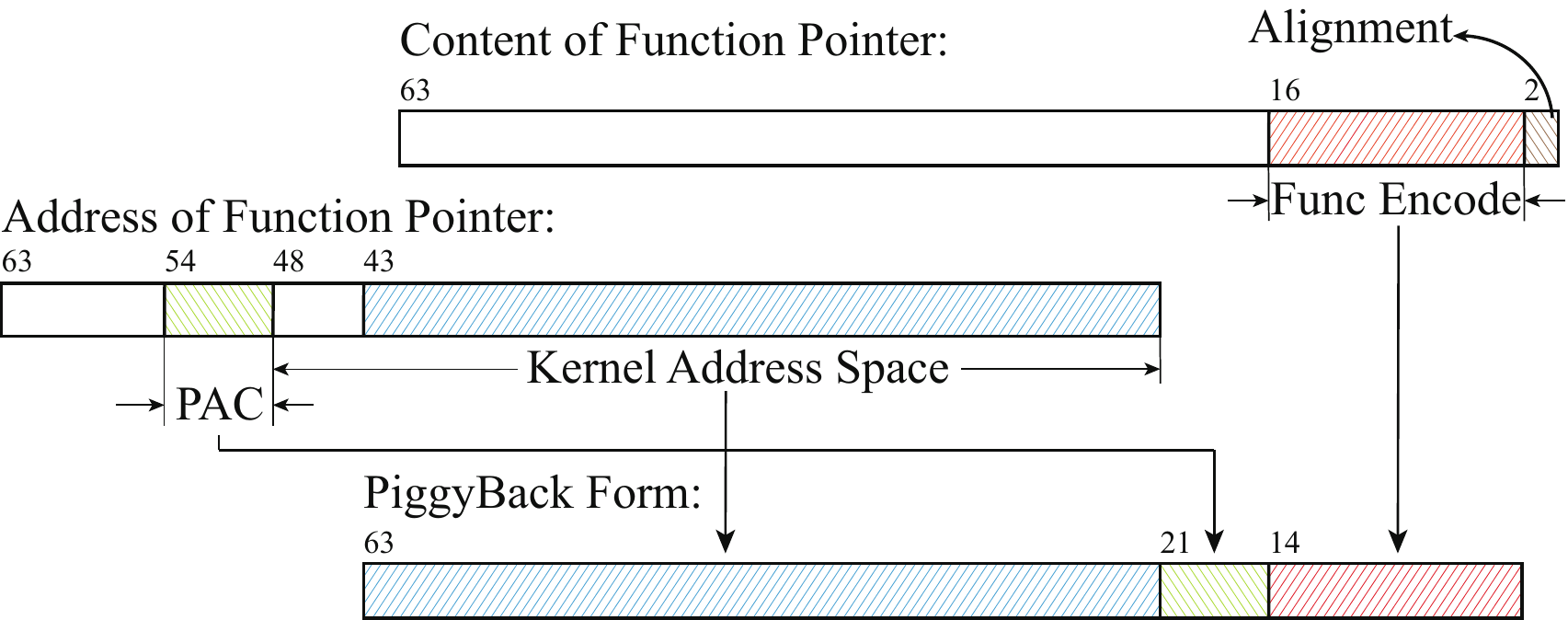}
    \caption{The pointer format with pointer-address piggyback. A piggyback pointer contains the function pointer value, the PAC, and the function pointer address.}
    \label{fig:piggyback}
    \vspace{-3ex}
\end{figure}
\subsection{Function Pointer Address Propagation} 
\label{sec:piggy}
Pointers with PAC should be authenticated at both the \textit{function pointer load} (loading a function pointer from memory to a register) and the \textit{function pointer branch} (jumping to a function pointer). In address-based PAC, authenticating at load site is straightforward as the pointer's address and value can obtain easily from the load instruction. 

However, for PAC authentication at branch (call) site, deciding the address of a function pointer becomes much harder. If the function pointer call is within the same function as the pointer loading, we can get the function pointer address by going through the use-def chain of the function pointer. 
However, in kernel, the gap between the loading and branching can be inter-procedure. For example, the kernel has hundreds of places that load a function pointer and pass its value as a parameter to a callee function, while the actual function pointer branch is in the callee function, as shown in \code{\_\_async\_schedule} in~\cref{fig:FPArg} and \code{do\_dentry\_open} function in~\cref{fig:load-branch-code}. In those cases, it is  impossible to specify the function pointer address at the callee site. One native solution can be changing the callee function by adding the address as an additional parameter. However, changing hundreds of calling functions is not practical.



In order to address the function pointer address propagation problem, we propose \textit{pointer-address piggyback}. The basic idea is to piggyback the address on the function pointer, so that the function pointer always carries its address. As a result, we can always get the function pointer's address whenever we use the function pointer. 
To achieve pointer-address piggyback, we encode the function pointer value so that the encoded function pointer contains the point-to value, the PAC, and the address, as shown in~\cref{fig:piggyback}. Our key observation is that for kernel with \code{defconf} configuration, the total number of address-taken functions is less than 10k, which can be encoded by 14 bits ($2^{14} == 16k$). Therefore, we can use 14 bits to index a function pointer value (the point-to address). PAC needs 7 bits, giving us 43 bits for encoding the address, as shown in~\cref{fig:piggyback}. As ARM/ARM64 are word (4 bytes) aligned, which means the 43 bits can be used to index 45 bits of virtual memory.

\algdef{SE}[DOWHILE]{Do}{doWhile}{\algorithmicdo}[1]{\algorithmicwhile\ #1}%
\begin{algorithm}
\caption{Precise function pointer identification}
\label{alg:dfa}
\begin{algorithmic}[1]
\Function{AnalyzeModule}{\textit{M}}
\State $\textit{S},\ \textit{S}_\textit{GI} \gets \Call{AnalyzeGlobalFP}{M}$\label{dfaline:2}
\State $\textit{STI} \gets \Call{AnalyzeStruct}{M, \textit{S}_\textit{GI}}$\label{dfaline:3}
\Do
\State $\textit{S'} \gets \textit{S}$
\ForAll {$\textit{F} \in \textit{M} $}
\State $\textit{S},\ \textit{STI} \gets \Call{AnalyzeFunction}{\textit{F}, \textit{S}, \textit{STI}}$
\EndFor
\doWhile{$\textit{S}\ne\textit{S'}$} %
\State \Return $\textit{S}$
\EndFunction
\State
\Function{AnalyzeFunction}{\textit{F}, \textit{S}, \textit{STI}}
\ForAll{$\textit{BB} \in \textit{F}$}\label{dfaline:14}
\ForAll{$\textit{I} \in \textit{BB}$}
\State $\textit{S},\ \textit{STI} \gets \Call{AnalyzeInst}{I, \textit{S}, \textit{STI}}$\label{dfaline:16}
\EndFor
\EndFor\label{dfaline:18}
\ForAll{$_\textit{rev}\textit{BB} \in \textit{F}$}
\ForAll{$_\textit{rev}\textit{I} \in \textit{BB}$}
\State $\textit{S},\ \textit{STI} \gets \Call{AnalyzeInst}{I, \textit{S}, \textit{STI}}$\label{dfaline:21}
\EndFor
\EndFor
\State \Return $\textit{S},\ \textit{STI}$
\EndFunction
\State
\Function{AnalyzeInst}{\textit{I}, \textit{S}, \textit{STI}}
\ForAll{$\textit{op} \in \textit{I}$}
\If{$\Call{IsFunctionPtr}{\textit{op}, \textit{STI}}$}
\State $\textit{S} \gets \textit{S} \cup \{op\} $\label{dfaline:30}
\EndIf
\EndFor
\State $\textit{S} \gets \Call{Propagate}{\textit{I}, \textit{S}}$
\State $\textit{STI} \gets \Call{Update}{\textit{S}, \textit{STI}}$
\State \Return $\textit{S},\ \textit{STI}$
\EndFunction
\end{algorithmic}
\end{algorithm}

\subsection{Identifying Function Pointers}
\label{sec:identFP}

To locate instructions to be instrumented, identifying function pointers among all the variables inside a module is required to be precise. Any mistaken instrumentation causes unexpected behaviors or even kernel panic. Note that only relying on function pointer type can hardly cover all of the function pointers, as some function pointers are typed as \textit{void*} or worse, 64-bit integer. Only when we combine program semantics on these variables, e.g., they become targets of indirect calls or they are assigned with function pointers, can we identify them as function pointers. We have also found that some fields inside a struct type are not function pointer type but contain function pointers, as shown in~\cref{fig:FPTypeChange}. Recording these fields can help us to discover more corner cases.

Following these insights, we have developed an intra-procedure and field-sensitive analysis method to precisely identify all the function pointers based on LLVM IR, whose details are given in \cref{alg:dfa}. Note that function pointer identification is totally different from function pointer alias (point-to) analysis. \name{} only requires to distinguish function pointers among all variables, while the later one tries to determine the point-to set of a function pointer.

Before diving into the details, we need to clarify the terms. \textit{M} denotes a module, \textit{F} denotes a function and \textit{I} denotes an instruction. Set \textit{S} in $\Call{AnalyzeModule}{}$ contains all the identified function pointers by our algorithm. We use \textit{function pointer field} to denote a field that contains a function pointer inside a struct. \textit{STI} stores all \textit{function pointer fields} to support our analysis. 

\subsubsection{Global Information Collection}
\label{sec:ident_gic}

The first step is to analyze data structure and statically initialized global variables, whose results can serve as basic information for subsequent function pointer identification and function pointer patching during kernel's early boot~\cref{sec:ptr-init}.

In $\Call{AnalyzeGlobalFP}{}$ at~\cref{dfaline:2}, we first walk through the initialization list of all global variables. The set of global variables which are initialized by function names are used to initialize \textit{S}, including those laying inside a struct. If they are part of a struct, the struct as well as their fields inside the struct are also stored into the set $\textit{S}_\textit{GI}$. 

The set $\textit{S}_\textit{GI}$ enables us to pick out initial \textit{function pointer fields}: if a field with its struct type in set $\textit{S}_\textit{GI}$, or if it is of function pointer type, it will be stored to \textit{STI}. This step is done at \cref{dfaline:3}. 
Each record of a \textit{function pointer field} consists the type name of a struct and a sequence of indices to reach this field. 
These struct names and indices can be duplicated because a struct can be nested in kernel, such as \code{task\_struct}. We organize \textit{STI} by a directed acyclic graph so that we can store all \textit{function pointer fields} with faster query speed and less memory overhead. A node in the graph represents a struct type that contains a \textit{function pointer field}, or just a basic type which may contain function pointers if it has no successor. An edge from node A to B indicates type B is included in type A.

By analyzing module global information, we get initialized \textit{S} and \textit{STI}, which hold basic information for analysis on each function. After finishing analysis on a function, \textit{S} and \textit{STI} are also updated by the analysis results. We analyse each function inside the module in one iteration and our algorithm will iterate continuously until \textit{S} no further changes.

\subsubsection{Per-Function Pointer Identification}

As described from \cref{dfaline:14} to \cref{dfaline:18}, rather than considering conditional or loop relationships among basic blocks, we linearly walk through each instruction inside a function because they do not affect the analysis result. More specifically, we focus on type information and its propagation across values, taking a loop just once or multiple times makes no difference on our analysis.

For each IR instruction, all its operands that can be identified as function pointers by global information are first added to \textit{S}, i.e., we check for each operand if it is of function pointer type, or comes from a \textit{function pointer field}. Then a propagation rule based on the kind of the IR instruction, namely the transfer function of the IR, is enforced to decide whether the rest operands and the instruction output should be added to \textit{S}. So far we have implemented transfer functions for seven kinds of IR instructions: \code{bitcast}, \code{icmp}, \code{phinode}, \code{store}, \code{load}, \code{getelementptr} and \code{call}, the details of which are further revealed in~\cref{sec:integration}. 

Note that our algorithm iterates twice----first in forward and then in backward directions inside a function as indicated at \cref{dfaline:16} and \cref{dfaline:21}. We do not simply iterate only once since an indirect \code{call} instruction does not propagate function-pointer attribute but generates the attribute for the called pointer. The function pointer operand added to set \textit{S} by an indirect call cannot propagate to the operands in the previously visited instructions. Therefore, we add an extra backward iteration to fully propagate function-pointer attribute. Intuitively, set \textit{S} will become stable after the two rounds as no more function pointer attribute is generated during the second iteration----it just propagates along the instructions.

\begin{figure}
    \centering
    \includegraphics[width=\linewidth]{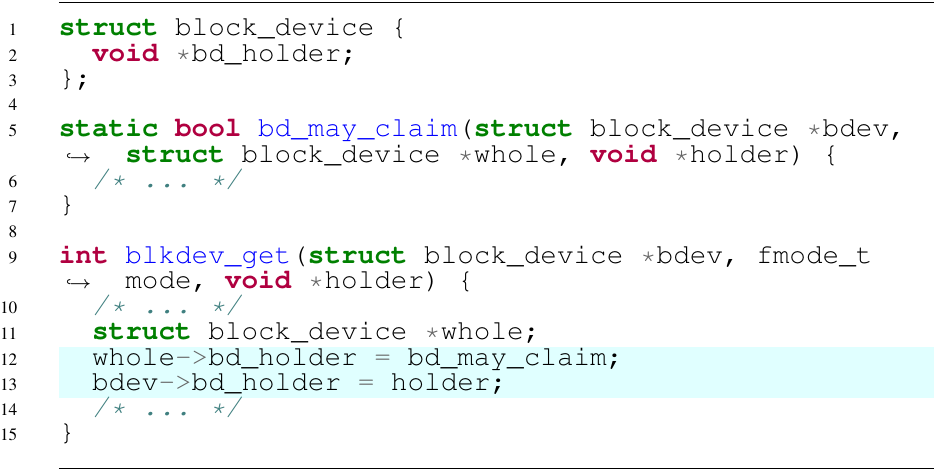}
    \caption{Even though both sides of the assignment at line 13 are \code{void*} type. \name{} can identify both of them as function pointers through field-sensitive analysis.}
    \label{fig:FPTypeChange}
    \vspace{-3ex}
\end{figure}

\subsubsection{Precision}
Our pointer identification algorithm is able to precisely identify most of function pointers inside the kernel, can eventually achieve 100\% after handling the corner cases in kernel. These corner cases and our measures will be later detailed in~\cref{sec:issues}. Generally, the precision of our algorithm comes mainly from three aspects: 

\textbf{Exclusiveness of function pointers}: we assume a function pointer is exclusive, i.e., it should always point to executable code after initialization and should not contain data of any other types at any time. In fact, this rule works in most times because mix use of function pointers and other data in one variable could be dangerous. For example, a variable that contains a data pointer could be called in this case, which leads to arbitrary code execution. The exclusiveness has made our analysis more precise because in our algorithm, a variable \textit{must} or \textit{must not} be a function pointer, rather than \textit{may} be.

\textbf{Semantic awareness}: Rather than depending only on type information, \name{} also utilizes program semantics to identify function pointers. Based on the semantics of the seven kinds of instructions we have modeled, \name{} are capable of identifying more function pointers.

\textbf{Field sensitivity}: \name{} takes advantage of field-sensitive analysis to achieve higher precision. 
\cref{fig:FPTypeChange} gives us a code snippet in kernel. At line 13, both \code{bdev->bd\_holder} and \code{holder} are of \code{void*} type and they are neither called or assigned with function pointers within \code{blkdev\_get}. Consequently, an algorithm with only semantic awareness will not identify them as function pointers.

However, \name{} records all the \textit{function pointer fields} to assist function pointer identification. Note that the same field \code{bd\_holder} in \code{block\_device} has been assigned a function \code{bd\_may\_claim}. So \name{} considers this field as a \code{function pointer field} and records it in \textit{STI}. When \name{} encounters the same field inside \code{bdev->bd\_holder}, it immediately identifies the variable as a function pointer. Then, \code{holder} is also identified as a function pointer from semantics of the assignment. 

The precision of our algorithm ensures that instrumentation instructions are inserted into correct locations. Inserted instructions will enforce function pointer integrity and return address integrity, which are introduced in the next two sections.

\begin{figure}[!t]
    \centering
    \begin{subfigure}{\linewidth}
        \includegraphics[width=\linewidth]{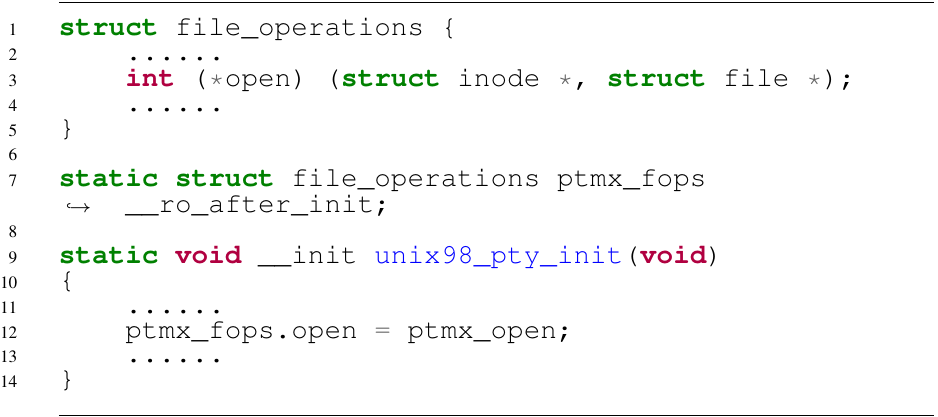}
        \caption{C code of \code{ptmx\_fops.open} assignment. Line 12 shows that \code{ptmx\_open} is assigned to the function pointer \code{open} inside the struct \code{ptmx\_fops}.}
        \label{fig:store-ptmx}
    \end{subfigure}\\
    \begin{subfigure}{\linewidth}
    \includegraphics[width=\linewidth]{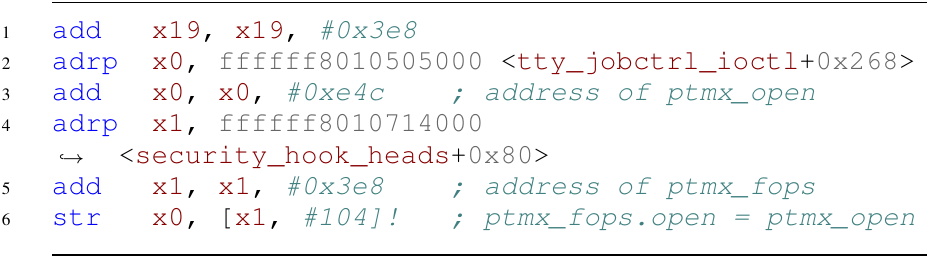}
    \caption{Assembly code of \code{ptmx\_fops.open} assignment without \name{} instrumentation. Line 6 shows the actually function pointer store operation, in which \code{x0} holds the function pointer value.}
    \label{fig:store-without}
	\end{subfigure}
	\begin{subfigure}{\linewidth}
    \includegraphics[width=\linewidth]{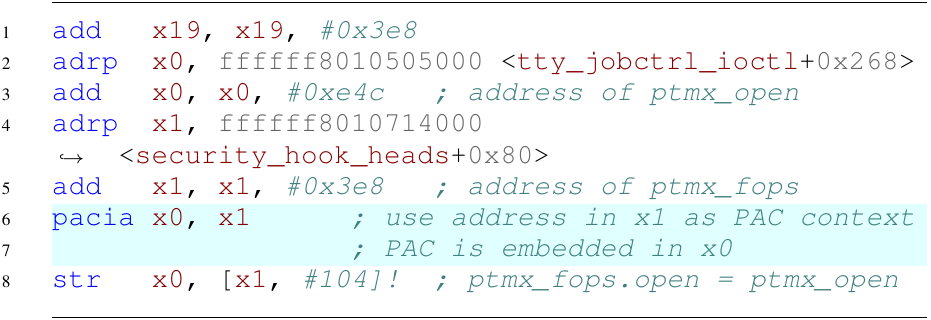}
    \caption{Assembly code of \code{ptmx\_fops.open} assignment with \name{} instrumentation. Line 6 shows the PAC generation instruction, in which \code{x0} holds the pointer value, \code{x1} holds the address of the pointer.}
    \label{fig:store-with}
	\end{subfigure}
    \caption{Function pointer store operation in \name{}.}
\label{fig:fp-store}
\vspace{-3ex}
\end{figure}

\subsection{PAC Instrumentation on Function Pointer Store/Load/Branch}
\label{sec:instrument}

\subsubsection{Generating PAC on Function Pointer Store}
\label{sec:fp-store}
All function pointers in memory should be protected by PAC to defeat attacker's corruption, thus they should be PACed before storing into memory. Function pointers are stored into memory under the following three cases: 1) Statically initialized function pointers which are loaded from binary to memory. 2) Dynamically assigned function pointers which are saved by store instructions. 3) Byte-object function pointers which are treated as a sequence of bytes and copied from another address by \code{memcpy} and \code{memmove}. 

Function pointers in the first case are PACed dynamically during kernel initialization after PAC keys setup. 
In fact, some dynamically assigned function pointers before PAC key initialization also need to be patched and we leave the details to~\cref{sec:ptr-init}. 

Most of kernel function pointer store falls into the second case. For this case, \name{} gets the function pointer and the address to be stored, PACs the function pointer before the storing instruction, as shown in~\cref{fig:fp-store}. Line 12 in \cref{fig:store-ptmx} shows that \code{ptmx\_open} is saved into function pointer \code{open} inside struct \code{ptmx\_fops}. Line 6 in~\cref{fig:store-without} shows the actual store instruction. \cref{fig:store-with} shows the assembly code after \name{} instrumentation. Line 7 shows the PAC generation instruction, in which \code{x0} holds pointer value while \code{x1} holds the pointer's address. Some already loaded function pointers may be in \textit{piggyback} form by our design, while others may be in normal form, e.g., immediate function addresses. \name{} is able to distinguish these two kinds of pointers as normal kernel function pointers have a fixed pattern in most times. But piggybacked pointers never follow this pattern. They are decoded to normal pointers before PACed.

The third case is more special as function pointers are decomposed to bytes and lose their type information during \code{memcpy} or \code{memmove}. We wrap these functions with \code{pac\_memcpy} and \code{pac\_memove} and replace them. Our wrapper functions check every byte of the \textbf{destination} object and use the pattern of PACed function pointers to match function pointers. We do not match on the source object as it could be overlapped with the destination and changed during \code{memmove}. Strict matching rules are adopted in matching.
Old PAC is stripped from a recognized function pointer, which is then patched with new PAC calculated by its new address as the context.

\begin{figure}[!t]
    \centering
    \begin{subfigure}{\linewidth}
        \includegraphics[width=\linewidth]{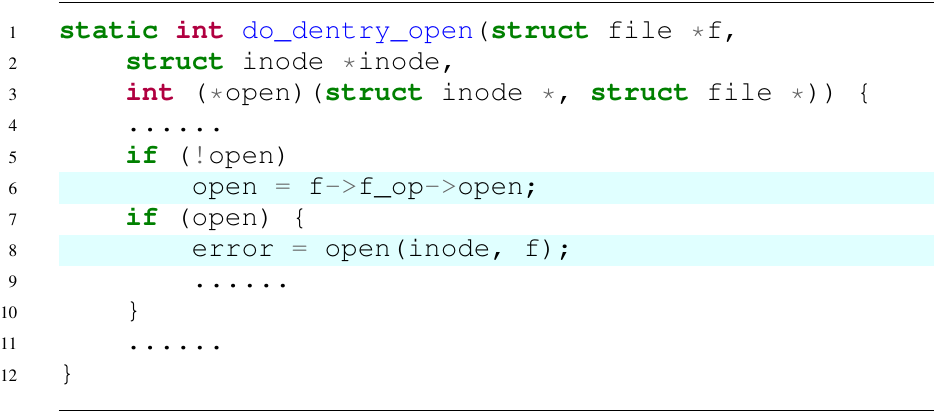}
        \caption{C code of load and branch operations. Line 6 shows a function pointer loads while Line 8 is a function pointer branch (function pointer call).}
        \label{fig:load-branch-code}
    \end{subfigure}\\
    \begin{subfigure}{\linewidth}
    \includegraphics[width=\linewidth]{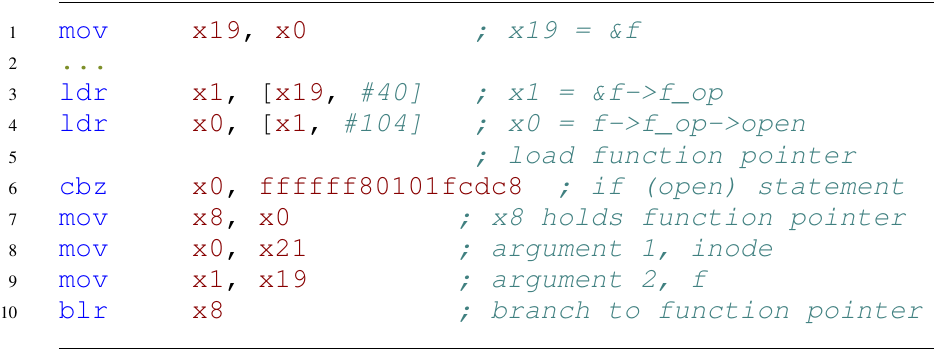}
    \caption{Assembly code of \code{ptmx\_fops.open} without \name{} instrumentation. Line 4 loads the function pointer to register \code{x0}. Line 7 moves function pointer to register \code{x8}. Line 10 is the function pointer branch, which calls function pointer in \code{x8}.}
    \label{fig:load-branch-without}
	\end{subfigure}
	\begin{subfigure}{\linewidth}
    \includegraphics[width=\linewidth]{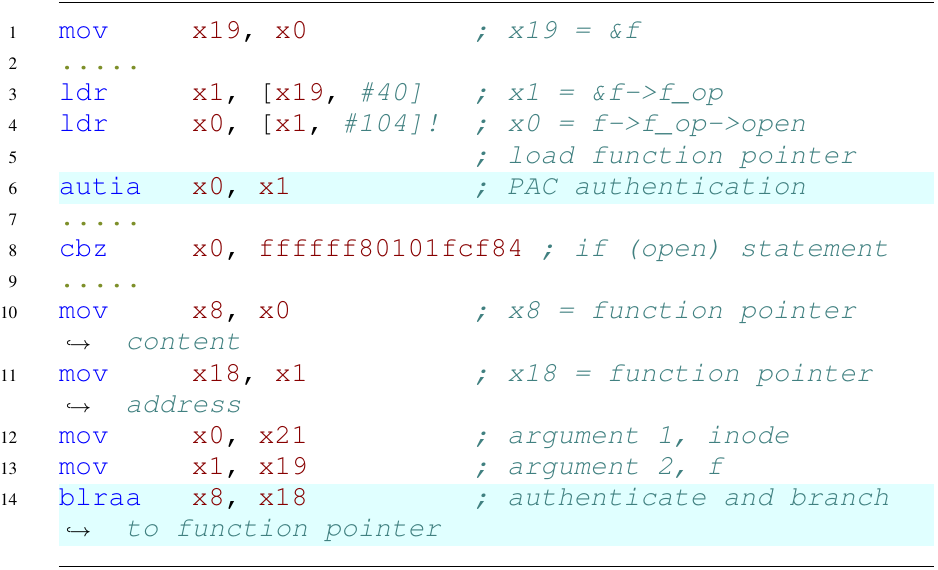}
    \caption{Assembly code of \code{ptmx\_fops.open} with \name{} instrumentation. \name{} adds Line 6 and Line 14. Line 6 is the PAC authentication code while \code{x0} holds the PACed pointer while \code{x1} holds its address. Line 18 is the authenticate and branch instruction, which authenticates the PACed pointer in \code{x8} first. \code{x18} holds the pointer address.}
    \label{fig:load-branch-with}
	\end{subfigure}
    \caption{Function pointer load and store operations in \name{}.}
\label{fig:load-branch-whole}
\vspace{-3ex}
\end{figure}

\subsubsection{Authenticating PAC on Function Pointer Load and Branch} 
\label{sec:fp-loadblr}

\name{} authenticates function pointers on both function pointer load and branch. Authenticating function pointers on branch instructions is easy to understand because it prevents function pointer branch instructions from being abused as JOP gadgets to jump to arbitrary addresses. 
However, function pointer loads also need to be authenticated for two reasons. First, the function pointer is loaded from memory, thus may be corrupted by the attacker, as the attacker has arbitrary memory write capability. Second, the function pointer is loaded into registers, thus it may propagate across different registers and used by other instructions, such the PAC generation instruction and store instruction. Without function pointer load authentication, the attacker can corrupt one function pointer in the memory to be a malicious code address. When the function pointer is loaded, a malicious code address can propagate to the PAC generation and store instructions. As a result, the PAC generation and store instructions can be abused by the attacker as a signing gadget. To break this signing gadget, we either authenticate function pointer before PAC generation instruction or authenticate on function point load. As the PAC generation instructions must accept un-PACed function pointers, we cannot authenticate them before PAC generation. Therefore, \name{} authenticates all function pointer loads, so that no illegal function pointer can sneak into registers. 

To authenticate function pointer load, \name{} instruments all loads with PAC authentication. Line 6 in \cref{fig:load-branch-code} shows a function pointer load, which is compiled into \code{ldr} instructions of Line 3-4 in~\cref{fig:load-branch-without}. Line 6 in~\cref{fig:load-branch-with} shows the \name{} instruments PAC authentication instruction that authenticates the PACed pointer in \code{x0} with \code{x1} holding the address as the context.



The other place to check PAC is at function pointer branch instruction (indirect call site). If a function pointer passes PAC check upon loading, it is transformed into piggyback form. A piggyback pointer is then extended to a PACed function pointer for the second PAC check at function pointer call site (function pointer branch).
As shown in~\cref{fig:load-branch-with}, at the function pointer call site, the original \code{blr} instruction is replaced by \code{blraa} in Line 14, which authenticates the pointer in \code{x8} first by using address in \code{x18} before branching.

\subsubsection{Function Pointer PAC Instruction Insertion}
\label{sec:fp-backend-insert}
To insert PAC generation and authentication code shown in~\cref{fig:store-with} and \cref{fig:load-branch-with}, \name{} performs instrumentation in both LLVM IR and LLVM backend, as shown by the ``Store/Load/Branch Instrument'' block and the ``FP PAC Instruction Insertion'' block in~\cref{fig:overview}.

In LLVM IR, \name{} first inserts IR instructions to indicate function pointer store, load, and branch respectively. The inserted calling instructions call different functions depending on the instrumented instructions, such as a call to \code{pac\_store} is inserted to replace function pointer store, \code{pac\_load} to replace function pointer load. For function pointer branch, things are a little bit different. We do not replace function pointer call, but add a \code{pac\_call} just before it. The parameters for \code{pac\_store} and \code{pac\_load} are the function pointer and its address, which can be fetched from operands of \code{store} and \code{load} instructions. For \code{pac\_call}, its parameter is the called function pointer, which is supposed to be a pointer in \textit{piggyback} form at run-time and can be decoded to a PACed pointer with its address. The inserted instructions will then be lowered from IR to LLVM machine instructions in the backend. 

Different from LLVM IR, LLVM machine instructions are closely related to the target architecture, including CPU-specific instructions. Therefore PAC-related instructions, which are only available on ARMv8.3 and newer ARMv8 architectures, are inserted in LLVM backend. Inserted machine instructions will replace the stub function calls we have added. 
For example, when \name{} backend encounters a call to \code{pac\_store}, it replaces it with a PAC-generating instruction followed by a store instruction storing the PACed pointer to the address just like what \cref{fig:store-with} shows. Note that the original function pointer and the address are already in register X0 and X1 as the function parameters of our stub calls. Dealing with \code{pac\_load} is just a similar process. However, stub calls to \code{pac\_call} is a little different. We must decode the input \textit{piggyback} pointer into a PACed pointer and its address, which are needed to be stored in two registers for \code{blraa}. Note that we cannot directly use X0 and X1 to save them because they are parameters passed into the indirect call. Instead, we save them to \textit{virtual register} provided by LLVM backend. The virtual registers will be mapped to registers that are not live at this point automatically by LLVM backend. \cref{fig:load-branch-with} displays the output of \name{} backend for indirect calls.


\begin{figure}[!t]
    \centering
    \begin{subfigure}{\linewidth}
        \includegraphics[width=\linewidth]{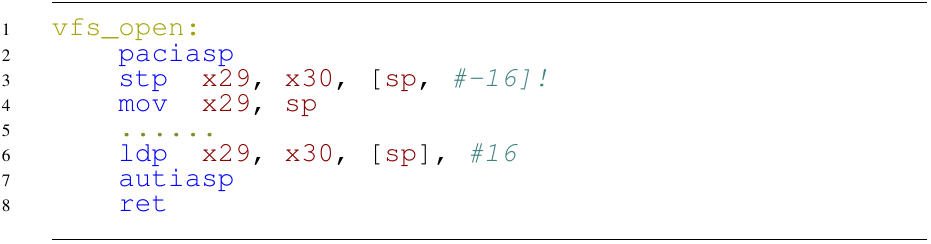}
        \caption{Existing design for protecting return address. Interrupt can happen between \code{autiasp} and \code{ret}.}
        \label{fig:qualcomm-ret}
    \end{subfigure}\\
    \begin{subfigure}{\linewidth}
    \includegraphics[width=\linewidth]{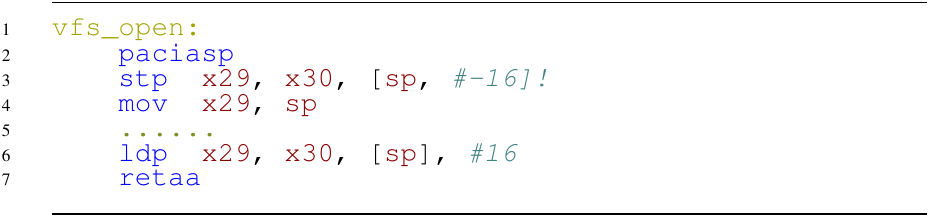}
    \caption{Return address protection in \name{}. Compared with existing design, \code{retaa} guarantees the atomicity.}
    \label{fig:packer-ret}
	\end{subfigure}
    \caption{Return address protection in \name{}.}
    \vspace{-3ex}
\end{figure}

\subsection{Return Address Protection}
\label{sec:ret-addr-prot}
The return address PAC generation and checking reuses the existing idea~\cite{qualcomm-ret-addr}: using the stack pointer as the context, generating the PAC for the return address register before pushing it to the stack and verify the PAC right after loading it from the stack, as shown in~\cref{fig:qualcomm-ret}. However, the existing design uses separate PAC authentication instruction \code{autiasp} and return instruction \code{ret}. The interrupt may happen in between, giving the attacker chances to launch \textit{time of check to time of use (TOCTTOU)} attack. For example, during the interrupt, the authentication of \code{x30} already passes, and its value will be saved on the interrupt stack. The attacker can overwrite \code{x30} value on stack and jump to an arbitrary place on return.

To address this problem, \name{} proposes to use the \code{retaa} instruction to authenticate and ret in one instruction. \name{} guarantees the atomicity and improves the security by eliminating the gaps between time of check and time of use.

Note that in kernel space, every thread has its dedicated kernel stack, while the stack pointer points to the stack frame within these per-thread kernel stacks. In other words, the stack pointer \code{sp} for different thread is guaranteed to be different, while using these stack pointer as the PAC context makes sure that the PAC cannot be replayed across different thread in kernel. More specifically, the stack pointer is per-thread, and per-stack-frame, making it virtually impossible to launch the pointer substitution attacks.

\subsection{Pointer Authentication Initialization}
\label{sec:pa-init}
Before utilizing pointer authentication (PA) to protect kernel function pointers and return addresses, we must first setup pointer authentication keys. Current Linux kernel only provides the PA for user space, not for kernel itself yet~\cite{kernel-pa, kernel-padoc}. Moreover, PA keys in Linux kernel are saved in kernel memory without any protection~\cite{kernel-pakey}, makes it vulnerable to arbitrary kernel memory read/write attacks.

To setup the PA environment for kernel, we set TBI and TBID bits in \code{tcr\_el1} as soon as \code{start\_kernel} is called to ensure PAC is 7 bits. Then we configure the \code{sctlr\_el1} to enable the IA and IB keys for PA. After that, \name{} needs to invoke kernel random number generation functions to generate 128 bits random numbers and set it into PA key registers.

Note that randomness functionality is not enabled at the very beginning of kernel boot. As a result, even the PA key registers are setup right after kernel randomness initialization, hundreds of functions that invoking function pointers are already executed before PA initialization. This will impose several challenges to \name{}. First, before PA key setups, a call to a function pointer may happen. As we mentioned before, a PAC check would occur at a function pointer callsite, the check will never pass without setting the keys. To avoid checking failure and crash, \name{} adds key setup check, if key has not setup, \name{} will not check PAC signature. Here \name{} only adds the check code to codes executed at kernel initialization process, and all those codes will be freed after kernel boot up, so this will not jeopardize the security of \name{}.

Second, hundreds of function get executed before PA initialization. These functions may have pointer load and store operations. Right after key setup, \name{} will patch the statically allocated function pointer with the correct PAC. However, the store operations happens before PA key ready stores the un-PACed pointer values. Therefore, \name{} must be able to trace all the store operations and patch all these locations. To address this problem, \name{} proposed to instrument all store operations. If PA is not ready, will allow the operation proceed, but will record the target address for later patch, details in~\cref{sec:ptr-init}. 

\subsection{Statically and Dynamically Initialized Function Pointer Patch}
\label{sec:ptr-init}
Same with userspace case~\cite{hans2019pac}, the statically allocated and initialized function pointers do not contain  PAC signature, as the pointer authentication key is not available at the compiling time.  Those function pointers need to be patched with proper PAC after we set PA key. However, different from user space, the kernel cannot rely on the loader, so it must figure out the addresses which need to be patched. 

To address this problem, during the compiling time, \name{} emits all addresses of the statically allocated functions. Especially for statically allocated kernel structures that contain function pointer fields, \name{} emits the address of the kernel structures and the offsets of the function pointer members.
For patching, \name{} maps this information to the actual pointer addresses during the kernel booting up. For each statically allocated and initialized function pointer variable or the function pointer member inside a statically allocated structure, \name{} first reads its value, calculates the PAC value, and writes back to the memory.  Note that this is all done after PA key is set.

As mentioned before, kernel randomization is not enabled at the very beginning. As a result, hundreds of functions get executed before the PA key is ready. These functions involve function pointers assignment. In other words, besides the statically initialized function pointers, \name{} also needs to patch function pointers that get initialized dynamically before PA key is ready. To do this, as all function pointer stores are already instrumented, \name{} checks the PA key status before the PA code generation and the store. If PA key is not ready, \name{} skips the PA code generation, just stores the raw pointer value. At the same time, \name{} will record the address of this function pointer. 
After PA key initialization, \name{} will come back to calculate the PA code for all recorded addresses, and update the pointer value with the correct PAC.

\section{Implementation}
\label{sec:imple}
In the implementation section, we first give out the environment settings we used for our implementation. Then we talk about the \name{} modification on compiler and the kernel. Finally, we present the details of the practical issues we encountered during our implementation.

\subsection{Environment Settings}
We have implemented \name{} on LLVM 10 and Linux kernel v5.0.1. \name{} kernel is built at optimization level O2. The kernel binary is running on ARMv8-A Fixed Virtual Platforms (FVP) based on Fast Model v11.7.30. FVP is a software simulator from ARM, which provides pointer authentication hardware simulations. FVP environment is set up on Ubuntu 18.04, running on Intel i7-7700. To boot up the kernel with \name{} on FVP, we also wrap a bootloader with \name{} kernel using boot-wrapper and build a minimal initram file system with buildroot.

\subsection{Compiler Modifications}
\label{sec:integration}
\name{} Clang/LLVM implementation contains 3 passes, about 2600 lines of code.
LLVM organises all its analysis and optimization functionalities in the unit of pass, therefore, both of our works on IR level and on backend are implemented in passes. LLVM passes are executed in a serial order and the positions of our passes in the order can affect the result. On IR level, IR output of the prior passes may be optimized by the optimization passes. Therefore, to avoid our inserted IR from being optimized, we put all our IR passes together at the end of all IR passes. On backend, LLVM machine code is lowered by passes. In this process, it loses high-level information, such as the type info and the virtual registers info, and gets closer to the real target assembly code. Some of our backend passes use virtual registers so they must run before the register allocation pass. The others are put to the end of the machine code passes.

For ease of use, We have also modified Clang/LLVM backend to support command line flags. One can pass \code{-mkfpi} flag to enable IR-level functionalities while pass \code{-mllvm -aarch64-enable-kfpi} flags to enable backend instrumentation. 

\subsubsection{IR-level Implementation}
We have added three passes to the LLVM passes to implement  function pointer identification~\cref{sec:identFP} and IR instruction instrumentation~\cref{sec:instrument}, namely \code{InitPass}, \code{MarkPass}, and \code{InstrumentPass}.

First, the \code{InitPass} analyzes global variable initialization and data structure that contains function pointers, and store the results for subsequent use. Note that it does not change IR code and will pass the IR to \code{MarkPass} as soon as it finishes its work. 

\code{MarkPass} is responsible for identifying all IR values which are function pointers, i.e., calculating the set \textit{S} in~\cref{alg:dfa}. Here we focus on implementation details about the transfer functions of the seven instructions:
\code{bitcast} transforms the type of an IR value and we put both the output and the input value into set \textit{S} if any of them is in \textit{S}. 
\code{icmp} compares two integers and decides the result according to the input condition. Therefore, either one of the two operands belonging to \textit{S} indicates the other should be also added to \textit{S}. 

\code{phinode} is an implementation of $\phi$ in static single assignment(SSA) form. It chooses one of inputs as its output according to its predecessor basic blocks. We just add all the values including the output to \textit{S} if one of these values is in \textit{S}. The strategy is based on our insight that in most cases, no value could be a function pointer in one condition but a data pointer or a normal integer in another. A special case is the union type which may violate our assumption, details in~\cref{sec:issues}.

\code{store} instruction saves the first operand into the memory pointed to by the second operand and has no output. If the first operand is a function pointer, then the second operand should be a pointer to a function pointer. 
Note that we have to distinguish function pointer and the pointer to a function pointer for \code{store} so we add an extra attribute \textit{level} to each element in set \textit{S}. Level 0 means the element is a function pointer, level 1 means a pointer to a function pointer and so on. Consequently, the second operand in \code{store} must be one level higher than the first operand. \code{load} instruction just does the opposite process of \code{store}. 

\code{getelementptr}, also known as GEP, takes a struct or array pointer and a sequence of indices as its inputs. It calculates the offset from the indices and outputs the pointer incremented by the offset. The instruction is usually used to index a field inside a struct or array. As we maintain all function pointer fields information including the struct types and indices in a DAG \textit{STI} (\cref{alg:dfa}), we can decide the result of a GEP is a level 1 function pointer if its inputs forms a path in the DAG.
Finally, \code{call} instructions just adds the called operand to \textit{S}.

IR code is not modified in \code{MarkPass} either and passed to \code{InstrumentPass}, where the instrumentation IR code is finally added according to \textit{S}. \code{InstrumentPass} also outputs the global variables initialized by function pointers with their offsets inside a struct to support function pointer  patching during early kernel boot.

\subsubsection{Backend Implementation}
We have added four backend passes to LLVM AArch64 backend, i.e., \code{VirtRegPass}, \code{BranchPass}, \code{RAPass} and \code{MemcpyPass}. \code{VirtRegPass} allocates virtual registers to save the PACed function pointer and its context needed by \code{blraa}. \code{BranchPass} changes all the \code{blr} to \code{blraa} using the two virtual registers as inputs. 
Both of the two passes must execute before register allocation. The other two are executed at the end of all the backend passes. 
\code{RAPass} is for return address protection, it first locates function frame setup and destroy and then inserts \code{paciasp} before frame setup and \code{autiasp} after frame destroy. 
\code{MemcpyPass} replaces all calls to \code{\_\_memcpy} and \code{\_\_memmove} to \code{pac\_memcpy} and \code{pac\_memmove}, respectively. Although this has been done once in IR, we do this just in case that a sequence of assignment on continuous memory is optimized to \code{\_\_memcpy} at backend, which actually happens at optimization level O2. 

\subsection{Kernel Patching}
\name{}'s kernel modification has about 600 lines of code, including initializing pointer authentication hardware and patching the un-PACed function pointers.
The PA initialization is mainly set up the registers to enable PA functionality. Moreover, it also calls the kernel random number generator to generate PA key. As a result, the PA can only be initialized after kernel enables the random number generation functionality. Therefore, PA initialization is done right after \code{add\_device\_randomness} in \code{start\_kernel}.

It is worth mentioning that PA initialization code contains the sensitive instructions that loading the PA key. To guarantee the security and remove all PA key manipulation related instructions, we mark all PA initialization code as init text, so that it will be freed by \code{free\_initmem} as soon as the kernel boot completes. As mentioned before, \name{} replaces all \code{blr} instructions to \code{blraa}. For \code{br} instructions, \name{} changes kernel build by adding \code{-fno-jump-tables} to instruct the compiler not use \code{br} instructions.


For function pointer patching, \name{} patches all statically initialized function pointers and the function pointers that get assignment before PA initialization.  To achieve this, \name{} uses a giant array to hold all the addresses of pointers to be patched. It also inserts code after PA initialization to generate PAC for each pointers. Again, the array is marked as init data, so that it will be freed after booting up to save the memory.

Note that \name{} is designed only for kernel code pointer protection, with only one PA key register is used, and leaving the other four PA key registers for user space PA protection. Therefore, \name{} is design with the consideration of user space PA compatibility.



\subsection{Practical Issues}
\label{sec:issues}
We encountered numerous practical issues during our implementation of \name{}. Due to the space limitation, here we only discuss several of them in detail.

\begin{figure}
    \centering
    \includegraphics[width=\linewidth]{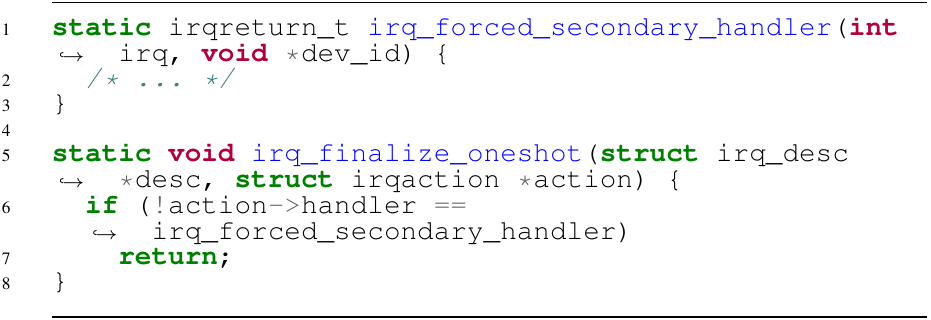}
    \caption{An example of function pointer comparison in kernel. Line 6 compares a function pointer with a function name.}
    \label{fig:FPCom}
\end{figure}

\subsubsection{Function Pointer Comparison}


In kernel, function pointers are usually used to compare with a function name directly, as shown by Line 6 in~\cref{fig:FPCom}. The comparison can be against a function name, which is the constant address of a function, or in some rare cases, against magic numbers like 1 or 2. In our implementation, the value loaded from the pointer would be transformed into the piggyback form, therefore its value will not match at every comparison. In those cases, we need to restore the function pointer after loading from the memory. The implementation of this part is pretty straightforward: our IR pass will traverse every \code{CmpInst} and check if the type of the operand is function pointer, if yes, replacing the operand with the restored value.

\begin{figure}
    \centering
    \includegraphics[width=\linewidth]{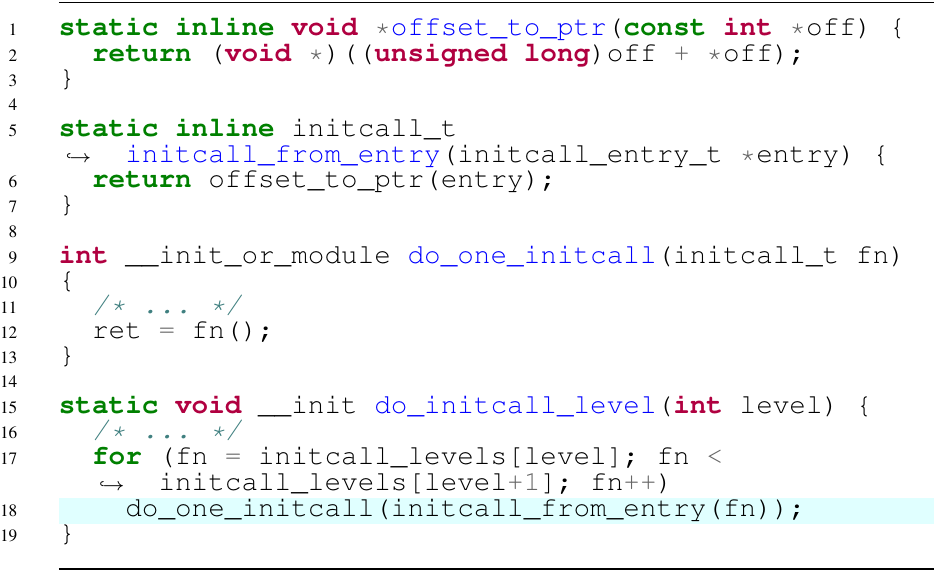}
    \caption{An example of function arithmetic in kernel. Line 17 contains function pointer increment \code{fn++}.}
    \label{fig:FPArithmetic}
    \vspace{-3ex}
\end{figure}

\subsubsection{Function Pointer Arithmetic}


Besides comparison, arithmetic on function pointers also exists in Linux kernel. For instance, \code{do\_initcall\_level} in Line 15 of~\cref{fig:FPArithmetic} passes function pointer \code{fn} to \code{do\_one\_initcall} while \code{fn} is calculated using the base address and the offset, as shown in Line 17. Fortunately, such case only appears once at kernel boot up stage. As we trust the kernel boot up, we consider the content in variable \code{initcall\_levels} as benign values, so in our implementation, after the function pointer is calculated, we generates the PAC using a constant context and the value of the pointer, so that it could pass the PAC authentication at \code{blraa}.

\begin{figure}
    \centering
    \includegraphics[width=\linewidth]{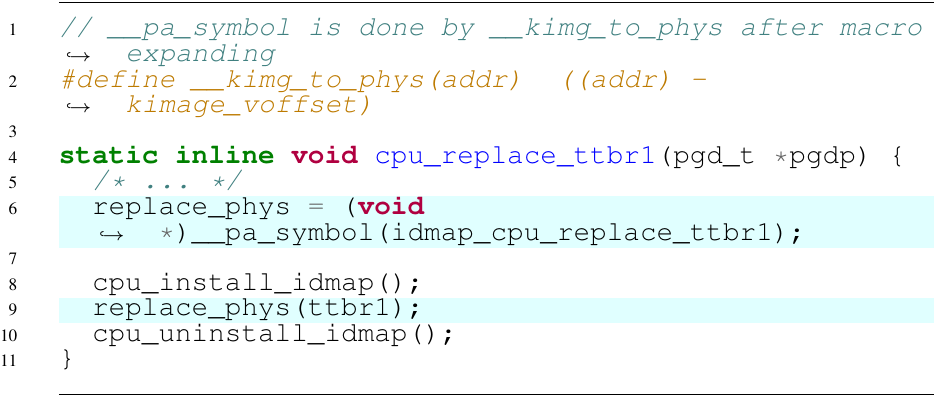}
    \caption{An example of physical address used as a function pointer. The function pointer \code{replace\_phys} gets the physical address at Line 6, is invoked at Line 9.}
    \label{fig:FPphys}
    \vspace{-3ex}
\end{figure}

\subsubsection{Function Pointer Holding Physical Address}
In Linux, for certain memory management unit (MMU) related function, the kernel will use its physical address directly, rather than virtual address (More precisely, for identical map, the virtual address and the physical address are the same).

As shown in~\cref{fig:FPphys}, function \code{idmap\_cpu\_replace\_ttbr1}'s physical address is assigned to function pointer \code{replace\_phys} at Line 7.
After that, the kernel turns off the memory management and directly branch to the physical address holding in \code{replace\_phys} at Line 11. Unfortunately, this breaks our rule that all operands of an indirect call should be a piggyback form pointer, and the system will go panic when the corresponding \code{blraa} is executed. As we mentioned in previous section, we add several instructions to change the pointer into piggyback form with a constant context before the indirect call, and this will not increase attack surface as the address is constant value from a \code{adrp} instruction, which cannot be changed by the attacker.




\begin{figure}
    \centering
    \includegraphics[width=\linewidth]{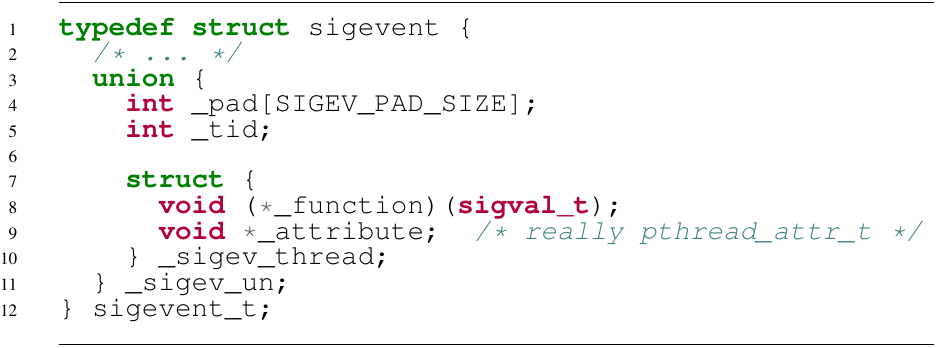}
    \caption{An exmaple of \code{union} that contains function pointer. \code{\_sigev\_un} is a union. Function pointer \code{\_function} is at Line 8.}
    \label{fig:FPUnion}
    \vspace{-3ex}
\end{figure}

\subsubsection{Function Pointer in Union}

Union type in kernel can contain a field that can be both function pointers and data such as integers, as shown in ~\cref{fig:FPUnion}. The data inside union variables are treated as function pointers or integers accordingly. This case breaks our assumption that a function pointer variable cannot contain data of other types.
As a result, \cref{alg:dfa} may mistake an integer for a function pointer. To address this problem, we use the alignment size of the field to distinguish whether a union field is a function pointer or not. Our key observation is that function pointer value is 64-bit, need to have 64-bit alignment. Therefore, if the program is using the field as an \code{int32}, the entry is considered to be used as data, either \code{\_pad} or \code{\_tid} in the figure; otherwise the type of the field is a function pointer, and we need to insert the \name{} code.

Note that the case is rare in Linux kernel and once we adopt the above rule, we can filter out all the troubles brought by union type.
\section{Evaluation}
\label{sec:eval}
In this section, we evaluate both the security and performance of \name{}. 

\subsection{Security Analysis}
For the security evaluation, we want to examine if \name{} can protect both function pointers and return addresses. Therefore, we first analyse the function pointer and return address coverage. We objdumped the generated vmlinux and extracted all the function pointer branch instructions and function return instructions.

For function pointer branch instructions, \name{} compiler component replaces all \code{blr} instructions in C code by \code{blraa} instruction during the compiling process. \name{} also manually changes the \code{blr} instructions in assembly code to \code{blraa}. As a result, 100\% of all function pointer branch are checked by \name{}. Compared with iOS kernel PA implementation which has \code{blr} residuals, \name{} contains no raw \code{blr} instructions, thus is more secure. Note that \code{blraa} does the PAC authentication and function pointer branch in one single instruction, giving the attacker no chance to launch time of check to time of use attacks.

For function pointer storing and loading, the pointer authentication hardware on ARMv8.3 does not provide the atomic instructions for the authenticate-store as well as the load-authenticate. Therefore, function pointer store and load are still vulnerable to time of check to time of use attacks. Here, we want to argue that this is a hardware limitation. Also, even though the store and load are not atomic, the final function pointer branch will be authenticated before branch atomically by \name{}, which can defeat any function pointer corruptions.


For the return address, we check all functions in vmlinux dump to examine that for all functions that pushing return address in prologue and popping return address in epilogue should be protected by \name{}. We go through the whole kernel dump using a script and our result shows that \name{} protects all return address pushing operations. For all return address pushing operations, \name{} inserts a PAC generation instruction. For all return address popping from stack, \name{} changes the return to \code{retaa} so that the return address will be authenticated before the actual return. Here, different from existing schemes of separating the PAC authentication instruction and the return instruction into two instruction~\cite{qualcomm-ret-addr, arm-pa}, \name{} uses a single instruction \code{retaa} to achieve the atomicity of both authentication and return. Therefore,  returns protections in \name{} is more secure by defeating time of check to time of use attacks.

\begin{figure}
    \centering
    \includegraphics[width=\linewidth, scale=1.00]{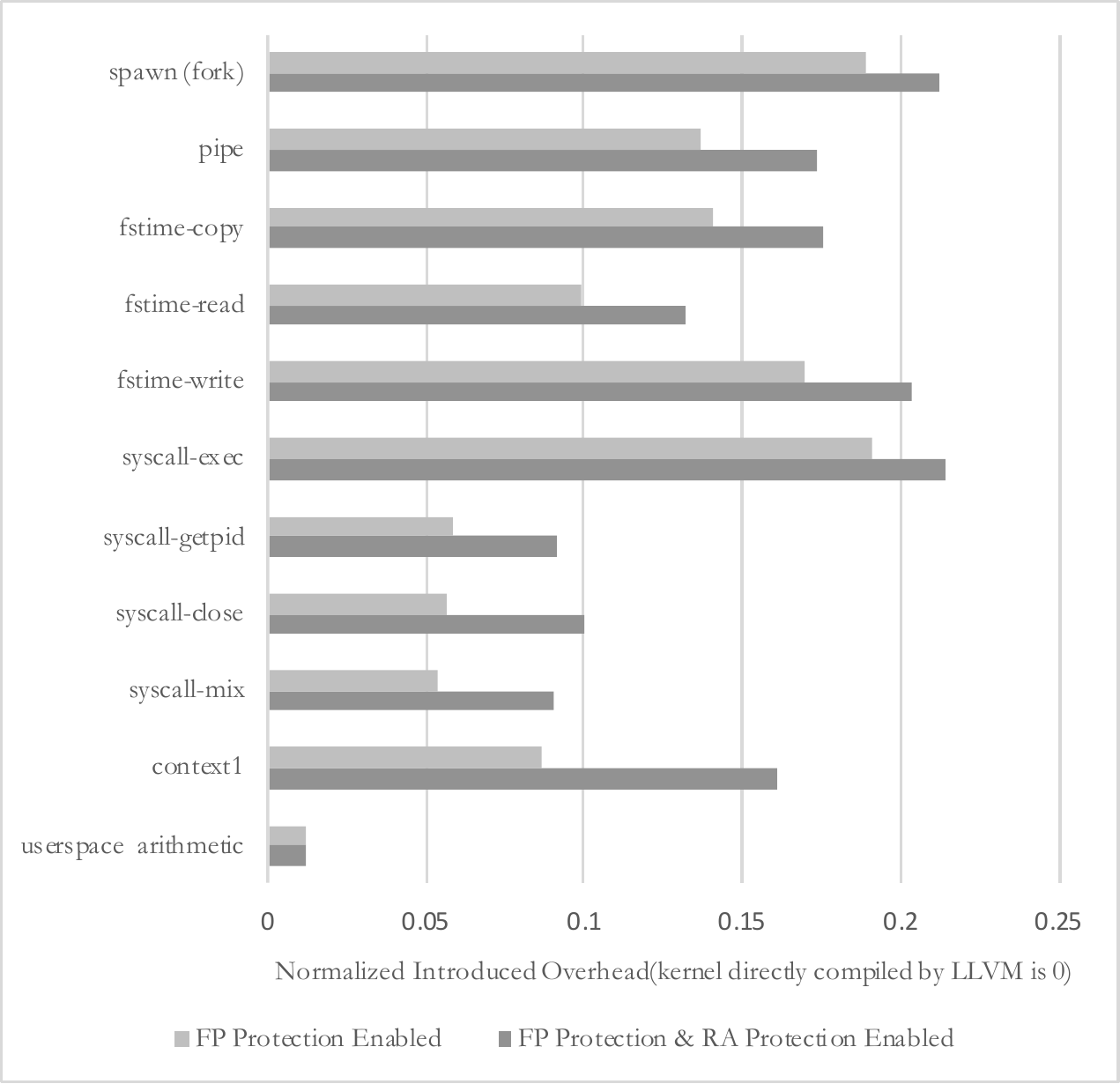}
    \caption{Performance evaluation of \name{} using Unixbench. Time overhead of the original kernel is normalized to 1.}
    \label{fig:eval}
    \vspace{-3ex}
\end{figure}

\subsection{Performance Analysis}
We choose Unixbench to evaluate \name{} performance. Unixbench is dedicated for unix-like systems and can measure performance of a system from different aspects. Three Linux kernels, which have the same version and configuration but different security level, are tested. One of them is compiled with original LLVM and is used as our baseline. The other two are both protected by \name{}, but one of them does not have return address protection. For each kernel, we have conducted all the tests listed in \cref{fig:eval} and these tests focus on critical system calls. Note that all the syscall-unrelated arithmetic tests like \textit{whetstone} and \textit{arithoh} have little connection with the performance of \name{} kernel, so their performance overhead are averaged and treated as the \textit{userspace arithmetic} test in \cref{fig:eval}.

Compared with the original kernel without \name{} protection, \name{} introduces around 10\%-20\% performance overhead. Complex syscalls like \code{fork} and \code{write} introduce more overhead because the number of stack frames and function pointer calls is larger. \name{} also introduces around 1\%-2\% overhead on userspace arithmetic because \name{} also protects kernel context switch and makes it a little bit slower. Note that performance overhead does not mean \name{} kernel is 10\%-20\% larger than the original kernel. In fact, \name{} image size is 7.0\% larger.

We believe that the performance overhead is not low, but reasonable and acceptable. It is reasonable because kernel itself is much more complex than most user application. It contains many function pointers and indirect calls. The calling stack in kernel is also badly nested. Besides, protection of context switch and indirect calls inside interruption handlers also add to our overhead. We argue that the result is also acceptable because it reflects the upper bound of \name{} overhead. In this evaluation, it is the pure syscall overhead that we have measured. A user application cannot call complex system calls like \code{fork} all the time. So for users of \name{} system,  the overhead is better than our result.

In our future work, we are planning more evaluation of \name{}, including the instruction count and performance overhead break down. We also plan to optimize the performance of \name{} based on the evaluation results.

\section{Related Work}
\label{sec:related}
There are variant CFI mechanisms to defend code reuse attacks. Among all defense mechanisms proposed against ROP, ASLR is widely used in modern operating systems, the address of the program will be randomized under such protections so that attackers can’t easily locate the gadgets. Beside ASLR, function pointer encryption is proposed with different encrypt methods~\cite{EncodePointer,Shuffler,PointGuard} to defend JOP attacks. In those methods, function pointers will be encrypted with a process/thread specific secret key and decrypted when being used.

CFI will compute a control-flow graph in advance to ensure the control transfers are within the pre-computed graph. And most CFI mechanisms can not protect both user programs and kernels due to the huge differences in between.  Moreover, as most software control-flow protection techniques suffer from high performance overhead, hardware-assists control flow protection mechanisms are proposed. These techniques~\cite{davi2014hardware,davi2015hafix, qiu2016physical,qiu2017control,zhang2018hcic,hans2019pac} leverage hardware feature or add extra hardware modules to realize protection operations, thus reduce the overhead. 


\subsection{User CFI}
\subsubsection{Software-Based CFI}
Compact Control Flow Integrity and Randomization (CCFIR)~\cite{zhang2013practical} protects both forward-edge and backward-edge control-flow integrity for binary executables. CCFIR implements a new code segment named Springboard which contains stubs of all indirect targets (i.e. function pointers and return addresses). CCFIR redirects all indirect \code{jump/call} instructions and \code{ret} instructions to jump to stubs in Springboard with specified policies. 
%
%
Bin-CFI~\cite{zhang2013control} provides control-flow integrity for COTS binaries. It uses a similar design to CCFIR instrument to enforce control-flow integrity. However, both CCFIR and bin-CFI are found to be insufficient~\cite{goktas2014out,davi2014stitching}. 


\subsubsection{Hardware-Assisted CFI}
Cryptographic CFI (CCFI)~\cite{mashtizadeh2015ccfi} employs cryptography mechanism to protect the control-flow integrity. Similar to \name{}, CCFI uses cryptographic MACs which is produced with AES. CCFI calculates and checks MACs of function pointers and return address when they are loaded. Thus, CCFI protects both forward-edge and backward-edge control flow integrity. And CCFI uses address as context to compute MAC which is the same design as \name{}. Opaque control-flow integrity (O-CFI)~\cite{mohan2015opaque} protects control-flow integrity by restricting indirect branch targets within an address bound. The address bound can be derived from source code or object code and can be randomized by code layout randomization~\cite{wartell2012binary}. When a program is loaded, O-CFI randomly selects a bound pair, which indicates the legal branch address region, from a bounds lookup table. And O-CFI needs the help of x86 segmentation selector to prevent accident leakage of the bounds lookup table. The overhead of O-CFI is 4.7\%.


As ROP attacks need to continuously execute several gadgets, KBouncer~\cite{pappas2012kbouncer}, as well as ROPecker~\cite{cheng2014ropecker}, use Last Branch Recording (LBR), which records last executed branches, to detect ROP attacks. And the latter one has an average overhead of 2.6\%. Since LBR only records last 16 branches, CFIMon~\cite{xia2012cfimon} uses branch trace store to break the limitation. The authors argue that CFIMon prevents both JOP and ROP attacks. And CFIMon has an overhead of 6.1\%, higher than KBouncer and ROPecker.

\subsection{Kernel CFI}
~\cite{li2018fine} and~\cite{ge2016fine} proposed fine-grained control-flow integrity solutions for kernel. Since computing kernel control flow graph is a tricky thing, they mainly focus on reducing the number of indirect control-flow targets in static analysis and both of them achieved more than 99\% of indirect control-flow targets. For enforcing control flow integrity~\cite{ge2016fine} uses restricted pointer indexing~\cite{wang2010hypersafe} to enforce kernel control-flow integrity while~\cite{li2018fine} uses a similar scheme named indexed hooks~\cite{li2011comprehensive}. 

KCoFI~\cite{criswell2014kcofi}, which extends secure virtual architecture (SVA)~\cite{criswell2007secure}, provides a coarse-grained but complete kernel control-flow integrity solution. KCoFI needs to recompile the whole operating system kernel into a virtual instruction set which benefits the security by ensuring security policies are not violated. And the formal model of KCoFI is only partial proved. Both KCoFI and \name{} need support of the compiler. However, KCoFI introduces a significant performance overhead due to the virtual instruction set while \name{} employs existing hardware feature and introduce a much lower overhead.


\section{Conclusion}
\label{sec:conclu}

This paper presents \name{}, which utilizes ARMv8.3 pointer authentication for kernel code pointer protection. In particular, \name{} generates PAC for every function pointer store and return address pushing to stack, and authenticates the PAC on every function pointer load and branch, and return address popping from stack. Moreover, to defeat pointer substitution attacks, we propose a novel address-based PAC generation based on the observation that all function pointers in kernel have a different virtual address, which can be used as the context to achieve unique PAC. To achieve address-based PAC, we design the pointer-address piggyback for address propagation. We also proposed new techniques for identifying function pointers, for pointer store, load and branch authentication and for handling statically initialized function pointers.

We have implemented a prototype of \name{} based on Clang/LLVM and Linux kernel. In our implementation, \name{} is able to protection 100\% of indirect call sites and return addresses. We further evaluated our implementation on ARM Fixed Virtual Platforms. For all eight tests, the performance overhead introduced by \name{} ranges from 15\% to 25\%. 

In our future work, we plan to test the performance of \name{} thoroughly, by using different techniques such as the instruction counting. Based on the evaluation results, we also plan to optimize \name{}.



\balance{
    \bibliography{refs}
}

\end{document}